\documentclass[%
aip,
rsi
 amsmath,amssymb,
 reprint,%
]{revtex4-1}

\usepackage{graphicx}
\usepackage{dcolumn}
\usepackage{bm}

\usepackage[utf8]{inputenc}
\usepackage[T1]{fontenc}
\usepackage{mathptmx}
\usepackage[switch, modulo]{lineno}

\usepackage{tabularx,booktabs}
\newcolumntype{Y}{>{\centering\arraybackslash}X} 
\usepackage{array}

\begin{document}

\preprint{AIP/123-QED}

\title{A high-Q microwave dielectric resonator for axion dark matter haloscopes}
\author{R.~Di Vora} \email{divora@pd.infn.it} \affiliation{INFN, Sezione di Padova, Padova, Italy} \affiliation{Dipartimento di Scienze Fisiche, della Terra e dell'Ambiente, Universi{\`a} di Siena, via Roma 56, 53100 Siena, Italy}
\author{D.~Alesini} \affiliation{INFN, Laboratori Nazionali di Frascati, Frascati, Roma, Italy}
\author{C.~Braggio} \email{caterina.braggio@unipd.it}\affiliation{Dipartimento di Fisica e Astronomia, Padova, Italy}\affiliation{INFN, Sezione di Padova, Padova, Italy} 
\author{G.~Carugno} \affiliation{INFN, Sezione di Padova, Padova, Italy}
\author{N.~Crescini} \altaffiliation{Present address: University Grenoble Alpes, CNRS, Grenoble INP, Institut N\'eel, 38000 Grenoble, France.}  \affiliation{INFN, Laboratori Nazionali di Legnaro, Legnaro, Padova, Italy}
\author{D.~D'Agostino} \affiliation{INFN, Sezione di Napoli, Napoli, Italy}
\author{D.~Di~Gioacchino} \affiliation{INFN, Laboratori Nazionali di Frascati, Frascati, Roma, Italy}
\author{P.~Falferi} \affiliation{Istituto di Fotonica e Nanotecnologie, CNR Fondazione Bruno Kessler, I-38123 Povo, Trento, Italy} \affiliation{INFN, TIFPA, Povo, Trento, Italy}
\author{U.~Gambardella} \affiliation{INFN, Sezione di Napoli, Napoli, Italy}
\author{C.~Gatti} \affiliation{INFN, Laboratori Nazionali di Frascati, Frascati, Roma, Italy}
\author{G.~Iannone} \affiliation{Dipartimento di Fisica E.R. Caianiello, Fisciano, Salerno, Italy} \affiliation{INFN, Sezione di Napoli, Napoli, Italy}
\author{C.~Ligi} \affiliation{INFN, Laboratori Nazionali di Frascati, Frascati, Roma, Italy}
\author{A.~Lombardi} \affiliation{INFN, Laboratori Nazionali di Legnaro, Legnaro, Padova, Italy}
\author{G.~Maccarrone} \affiliation{INFN, Laboratori Nazionali di Frascati, Frascati, Roma, Italy}
\author{A.~Ortolan} \affiliation{INFN, Laboratori Nazionali di Legnaro, Legnaro, Padova, Italy}
\author{R.~Pengo} \affiliation{INFN, Laboratori Nazionali di Legnaro, Legnaro, Padova, Italy}
\author{A.~Rettaroli} \affiliation{INFN, Laboratori Nazionali di Frascati, Frascati, Roma, Italy} \affiliation{Dipartimento di Matematica e Fisica, Universit{\`a} di Roma Tre, Roma, Italy}
\author{G.~Ruoso} \affiliation{INFN, Laboratori Nazionali di Legnaro, Legnaro, Padova, Italy}
\author{L.~Taffarello} \affiliation{INFN, Sezione di Padova, Padova, Italy}
\author{S.~Tocci} \affiliation{INFN, Laboratori Nazionali di Frascati, Frascati, Roma, Italy}

\date{\today}
             
\begin{abstract}
The frequency band 1-15 GHz provides exciting prospects for resonant axion haloscopes as indicated by cosmological and astrophysical arguments. Among the challenges currently addressed to reach the required sensitivity, the development of high quality factor cavities that tolerate multi-Tesla fields plays a central role.  
 We report a 3D resonator based on a right circular copper cavity with hollow cylinders that confine higher order modes around the cylinder axis. Its effective volume at 10.3\,GHz is $3.4 \cdot 10^{-2}$ liters, and under an 8\,T-field we measured an internal quality factor of more than 9 millions.
These parameters demonstrate the potential of this unique resonator to probe galactic dark matter axion at remarkable scan rates of 15\,MHz/day when the cavity is readout by a quantum-limited receiver.

\end{abstract}

\maketitle

\section{\label{sec:intro}Introduction}

According to cosmological models, galaxies are surrounded by dark matter (DM) halos that extend far beyond the visible structures. Though DM halos cannot be observed directly, their existence is inferred through observations of their gravitational effects, and play a central role in our understanding of the Universe evolution \cite{Bertone:2018}. The nature of particle dark matter is unknown, and this is one of the most compelling problems in fundamental physics.\\
Unrelated to this open problem, a hypothetical particle named axion was introduced in the late 80s to answer a fundamental question in particle physics, namely why in the strong sector the charge-parity symmetry is not violated \cite{Peccei:1977,Peccei:1977ur}. However, physicists later realized that sufficiently light axions could also solve the DM problem \cite{Ipser:1983mw}.    
As for detectability, their dominant interaction is via two-photon coupling in the Lagrangian 
\begin{equation}
  \mathcal{L}=g_{a\gamma\gamma}\,a\,\mathbf{E}\cdot \mathbf{B}\,, 
\end{equation}\label{eq:1}
where $g_{a\gamma\gamma}$ is the  axion-photon coupling constant, proportional to the axion mass $m_a$, $a$ is the axion field, $\mathbf{E}$ and $\mathbf{B}$ are the local electric and magnetic field vectors, respectively. The related process lifetime exceeds the age of the Universe, but P. Sikivie early in the 80's showed that if the field $\mathbf{B}$ in eq.\,\ref{eq:1} was provided by a laboratory magnet, it would have been possible to probe the existence of this dark matter candidate using microwave cavities permeated by multi-Tesla fields \cite{Sikivie:1983ip}. This is the basic principle of the so-called haloscope, that detects axions through their conversion to photons in a microwave cavity using the inverse Primakoff effect and a large magnetic field \cite{Sikivie:2021,IRASTORZA201889}.    

As implicit in eq.\,\ref{eq:1}, the axion to photon conversion probability in a cavity is proportional to $\int_V\mathbf{E}_{mnl} \cdot \mathbf{B}_{0}\,dV$, where $\mathbf{E}_{mnl}$ is the electric field of the resonant mode, $\mathbf{B}_{0}$ is the applied magnetic field  and $V$ is the ca\-vi\-ty volume. Thus if the cavity frequency $\omega_c$ matches the axion frequency $\omega_a=m_ac^2/\hbar $, the signal power in the cavity is given by
\begin{equation}\label{eq:2}
P_{a} = g_{a \gamma \gamma}^2 \frac{\rho_a}{m_a^2} \omega_c B_{0}^2 C_{mnl} V \frac{Q_c Q_a}{Q_c+Q_a} \; \, ,
\end{equation}


where $C_{mnl}$ is the form factor that accounts for the spatial overlap between the external magnetic field and a given cavity mode. If $Q_0$ is the unloaded quality factor of the mode, $Q_c$ in eq.\,\ref{eq:2} is given by $Q_c=Q_0/(1+\beta)$, with  $\beta$ coupling coefficient of the receiver antenna to the axion-sensitive cavity mode. 

As the required intense magnetic field values are available in solenoidal configurations, the right circular cylindrical cavity is typically the chosen geometry to maximize $C_{mnl}$

\begin{displaymath}
C_{mnl} = \frac{ \, |\int{d^3x\, \mathbf{B} \cdot \mathbf{E}_{mnl}(\mathbf{x})\;}|^2}{B_{0}^2 V \int{d^3x\,\epsilon_r(\mathbf{x}) \,|\mathbf{E}_{mnl}(\mathbf{x})|^2} \; } \; \, ,
\end{displaymath}
where $\epsilon_r$ is the relative dielectric constant inside the cavity volume.
Pertinent to the cavity resonator is also the last ratio in eq.\,\ref{eq:2}, which expresses the dependence of the conversion power from the reduced quality factor of the system \cite{Kim:2020kfo} $Q_a^{-1}+Q_c^{-1}$, with $Q_a^{-1}\simeq10^{-6}$ axion signal relative linewidth\cite{IRASTORZA201889}. 
When no signature of the axion is observed, exclusion limits are reported at a certain confidence level, assuming that axions saturate the dark matter density $\rho_a=\rho_{\rm DM}$, whose value is typically taken as $\rho_{\rm DM}=0.45\,$\,GeV/cm$^{3}$ [\onlinecite{IRASTORZA201889}]. 
For a typical haloscope experiment the power given by eq.\,\ref{eq:2} is in the range from $10^{-23}$ to $10^{-25}$\,W, several orders of magnitude below  the electronic noise power $P_N$ in state-of-the-art cryogenic receivers. 
For instance, the QUAX$-a\gamma$ haloscope \cite{Alesini:2021} recently set a 90\% confidence level limit to the axion-photon coupling $g_{a \gamma \gamma} < 0.766 \cdot 10^{-13}$\,GeV$^{-1}$ at $m_a\simeq 43\,\mu$eV ($\omega_c=2\pi\, 10.4\,$GHz), using a cavity with $V=80.56$\,cm$^3$, $C_{mnl}=0.69$, unloaded quality factor $Q_0=72000$, coupling coefficient $\beta=1$. Altogether, the expected power for these parameters at the benchmark for KSVZ models \cite{Kim:1979,SHIFMAN1980493} is $P_{a}= 3.8\times 10^{-24}$\,W.
Under the assumption that the noise power follows a Gaussian distribution, as is the case for Johnson-Nyquist noise, 
experiments rely on the radiometer equation \cite{Dicke:1946} $\Sigma=P_a\sqrt{t\Delta \nu}/P_N$ to set the integration time $t$ to satisfy $\Sigma\simeq 2$ over a bandwidth $\Delta \nu$.   

The Axion Dark Matter eXperiment (ADMX), based on a 136\,\,l-volume cylindrical cavity, has excluded axions with masses $2.7\,\mu$eV$-4.2\,\mu$eV ($650$\,MHz$-1025$\,MHz) \cite{Braine:2020,admxcollab2021} down to benchmark DFSZ axion models \cite{DINE1981199}. However, existing astrophysical and cosmological constraints within axion parameter space motivate the construction of new detectors to search for heavier axions\cite{IRASTORZA201889}, in the two decade window $10\,\mu$eV$-1$\,meV. 
The next generation of cavity experiments in the frequency range $(1-15)\,$GHz address challenges related to sensitivity loss due to the reduction in volume \cite{Stern:2015}.
In addition, the minimum detectable electromagnetic field in a cavity with a low noise linear amplifier is set by the standard quantum limit (SQL), an irreducible noise source that increases with frequency \cite{Lamoreaux:2013, Caves:1982}.  
Quantum-limited Josephson parametric amplifiers have been employed as preamplifiers in Dicke heterodyne receivers to readout haloscope cavities, as in ADMX and HAYSTAC experiments \cite{Braine:2020,Zhong:2018}. The latter has recently circumvented the SQL with an electronic readout involving two JPAs to squeeze the receiver noise and improve the search speed by a factor of 2 \cite{Backes:2021}. Both the QUAX$a \gamma$ \cite{Alesini:2021} and QUAX$a e$ \cite{Crescini:2020} haloscopes operated a JPA at approximately the SQL (system noise temperature of 0.9\,K at $\sim10$\,GHz), while in their future scientific runs they will take advantage of the large bandwidth of 
traveling wave parametric amplifiers \cite{Planat:2020, Ranadive:2021}.

\begin{figure*}[ht!]
\includegraphics[width=6.5in]{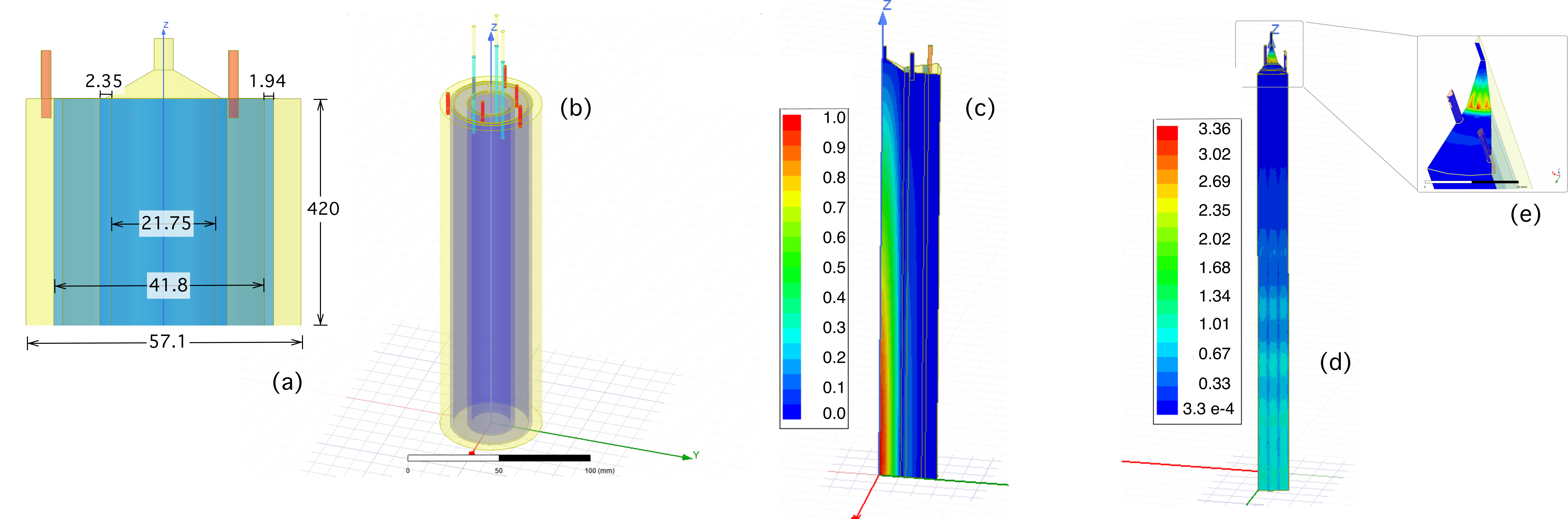}
\caption{\label{fig:1}  (a) Cavity model (not in scale) and relevant dimensions. Teflon pins are shown in orange. (b) 3D cavity model (in scale, half length) with copper external cylinder (light yellow), sapphire tubes (violet), teflon pins (red) and sapphire rods employed for tuning the cavity frequency (light blue). (c) Electric field profile of the TM$_{030}$ mode and (d) surface loss calculated via FEM simulations for a fraction of the overall cavity volume. (e) Surface loss in the cone structure.}
\end{figure*}

The optimized dielectric resonator presented in this work, designed to resonate around 10\,GHz with $\sim 500\,$MHz tuning range, addresses the high frequency challenge in haloscope detection by exploiting the dielectric resonator concept \cite{Morris:1984nu,McAllister:2018,Kim:2019asb,Alesini:2020vwh,Quiskamp:2020}. In this approach, the field profiles of higher-order resonant modes are properly modified by introducing dielectric structures that shape the cavity fields profiles, reducing the magnetic field amplitude on copper walls. This in turn helps to reduce dissipation and obtain higher quality factors. In addition, the dielectric material can suppress the out-of-phase components that negatively impact $C_{mnl}$ when higher order modes are considered. 
The present dielectric resonator, shown in Fig.\,\ref{fig:1}, has been improved compared to our previous realization \cite{Alesini:2020vwh} and we report unprecedentedly high quality factor values ever measured for a microwave resonator in multi-tesla fields.

\section{haloscope cavity design}

To optimize cavity design, we consider the cavity-based figure of merit $F=C_{mnl}^2V^2Q$. In fact, the appropriate expression to consider is that of the scan rate\cite{Stern:2015}, i.e. the speed at which haloscopes probe their targeted portions of axion parameter space for a fixed $g_{a\gamma\gamma}$ sensitivity

\begin{equation}\label{eq:3}
\frac{df}{dt} \approx \frac{g_{a \gamma \gamma}^4}{\Sigma^2} \frac{\rho_a^2}{m_a^2} \frac{B_{0}^4\, C_{mnl}^2 V^2}{k_B^2 T_s^2}\left(\frac{\beta }{1+\beta}\right)^2 \frac{Q_c Q_a^2}{Q_c+Q_a} \; \,,
\end{equation}
where $T_s$ is the system noise temperature \cite{Lamoreaux:2013}. 
Note that this expression gives the nominal scan rate, and an efficiency factor should be added to account for dead time during data runs.
 RF cavities host transverse electric (TE$_{mnl}$) modes, transverse magnetic (TM$_{mnl}$) modes, and transverse electromagnetic (TEM$_{mnl}$) modes. If  $\mathbf{B_0}=B_0\mathbf{z}$, the TE and TEM modes both lack axial E field components, resulting in null form factors. The modes that have non-zero form factors, and so of interest for axion detection, are the TM modes. For the TM$_{010}$ mode the form factor is $C_{010}=0.69$, while it is significantly smaller in higher order modes (C$_{020}=0.13$ and C$_{030}=0.05$).
 
An haloscope cavity must be operated under multi-tesla magnetic fields, that prevent the utilization of bulk niobium to reach high quality factors as is typically done in particle acceleration \cite{Padamsee:2014} and the conventional approach has been to use copper right circular cylinder cavities.
A most promising technology to reach high Q resonators for haloscope search is based on utilization of films of high-magnetic-field tolerant, type II superconductors such as Nb-Ti, NbSn3, or ReBCO. 
Copper cavities sputtered with type II superconductors as NbTi and NbTiN films have been investigated at liquid helium temperature under DC magnetic fields up to 6\,T, and the results of this search show that almost a factor 6 is lost at 6\,T starting from $Q_0=1.4\times 10^6$ measured at null field with a cavity resonant at 14.4\,GHz, and rapidly decreasing for greater fields  \cite{DiGio:2019}. Nonetheless, the $Q_0B^2$ factor for this cavity at $B = 2$\,T is still 5 times higher than in Cu, and it has been used to probe axions\cite{Alesini:2019ajt}, setting the limit $g_{a\gamma\gamma}<1.03\times 10^{-12}$\,GeV$^{-1}$ for about $37\,\mu$eV mass (9.08\,GHz). 
 Low loss in 8\,T magnetic field has also been accomplished with a biaxially textured YBCO cavity resonant at 6.9\,GHz, with $Q_0\sim 3\times 10^5$ [\onlinecite{Ahn:2021fgb}].

A strategy to probe for heavier axions is to increase $V$ by employing multiple cell cavities to efficiently exploit the volume available in the bore of a superconducting solenoid \cite{PhysRevLett.125.221302}, or to use long filter-based rectangular cavities when big volume dipolar magnets are available, as those that have been dismissed by LHC \cite{CAST:2021add}.  

The highest frequency pathfinder run has been conducted using a copper cavity resonant at $26\,$GHz \cite{MCALLISTER201767}, even though in the 10-100\,GHz frequency range the main experimental thrust is driven by the MADMAX project \cite{madmax:2020}, in which a system of layers with alternating dielectric constants is used to obtain an enhanced axion-photon conversion. 
 In this range a new class of thin-shell cavities has been recently proposed \cite{Kuo:2020llc}. 
 
 The dielectric cavity idea is based on dielectric materials properly placed inside traditional cylindrical resonant cavities, operated in TM modes of higher order \cite{Morris:1984nu}.   
Proof-of-concept experiments, including demonstration of a tuning mechanism, have been reported for a prototype Teflon-wedge dielectric cavity at room temperature, following accurate design based on results from finite-element modeling \cite{Quiskamp:2020}. Highly pure aluminum oxide (99.7\% Al$_2$O$_3$) was chosen as the dielectric material in a work by a different group, that investigated at 4\,K the mode mixing of the TM$_{030}$ mode with other cavity modes while tuning its frequency in the range (7.02-7.32)\,GHz \cite{Kim:2019asb}.
 
Frequency tuning mechanisms in conventional experiments are based on transverse symmetry breaking by displacing one \cite{Stern:2015} or multiple \cite{Simanovskaia:2021} metallic cylindrical rods. This can be performed by means of piezo-actuators or cryogenic motors. To demonstrate the resonator's effective tunability, careful comparison of simulation results and in situ tests are needed to address the problem of mode mixing, that takes place when other cavity modes cross with the axion-sensitive mode.  

The proposed tuning methods for dielectric resonators encompass splitting the dielectric hollow cylinder vertically into two halves and moving them apart along the radial direction \cite{Alesini:2020vwh,Kim:2019asb}, or breaking the symmetry along the cavity axis by spitting the dielectric in the longitudinal direction \cite{McAllister:2017ern}. 

In section\,\ref{tuning} we demonstrate a fine tuning mechanism based on dielectric rods, devised for preliminary tests of scientific runs. There is in fact much to be understood in the regime $Q_c>Q_a$, that this cavity allows to probe for the first time.  
The $\sim$\,1\,kHz-resolution, $\sim$\,1 MHz-maximum range 
tuning we obtained reveal new challenges for data analysis that will be reported in a forthcoming work.

\section{\label{sec:simu} FEM electromagnetic simulations and methodology}

In our previous work \cite{Alesini:2020vwh} we reported tests conducted with a dielectric resonator composed of a couple of concentric hollow sapphire cylinders centered about the axis of a cylindrical copper cavity (see Fig.\,\ref{fig:1}). In this type of cavity higher order modes are exploited, and we focused on the TM$_{030}$ mode for axion detection.
The cavity was designed using finite element methods (FEM) so that the product $CV$  was of the same order of magnitude for the TM$_{030}$ mode in the dielectric cavity and the TM$_{010}$ mode of a cylindrical copper cavity with same frequency and length.
The realized cavity presented some practical issues, as the cylinders were not properly blocked, and therefore the cavity parameters were not reproducible at every cooling run.
Most importantly, interference with a number of spurious modes was experimentally observed, that negatively impacted on the expected quality factor of the TM$_{030}$ mode. Comparison of the results of an improved finite-element analysis of the adopted solution with in-situ tests, has shown that these additional modes arise from the circular grooves made in the copper endplates to hold the dielectric shells in place. 
Another important issue in the previous realization was related to the irregular shape of the sapphire tubes we used. 

The cavity presented in this work has been re-designed to allow for centering of the dielectric shells axis to the copper cavity axis at cryogenic temperature. 
In the present configuration, the sapphire tubes are held in place by teflon pins as shown in Fig.\,\ref{fig:1}\,(a) and\,(b), without the need of grooves in the copper endplates.  
The cylindrical body is 420\,mm long and has 57 mm-diameter, while inner/external sapphire tubes have 21.75\,mm/41.8\,mm internal diameter and thickness 2.35\,mm/1.94\,mm. The cavity volume $V$ is then 1.0776\,l. 
In addition, the sapphire tubes we used for the present dielectric resonator have been shaped with mechanical tolerances chosen on the basis of FEM simulations. 
For instance, the required tolerances for the external/internal diameter of the internal cylinder were of $\pm 0.05$\,mm/$\pm 0.15$\,mm. Tolerances on concentricity, linearity and coaxiality have also been given. 
Confinement of the electric field around the cavity axis, that is crucial to suppress dissipation in the cavity walls, is visualized in Fig.\,\ref{fig:1}\,(c). 

The crystal tubes are grown using the Stephanov method by Rostox-N Ltd (Russia) \cite{Rostox}. Pure sapphire raw materials (at least 99.99\% Al$_2$O$_3$) are melted in a crucible made of Molybdenum (2050\,$^{\circ}$C temperature) and the melt is then pulled through a ring-shaped Molybdenum component to obtain a sapphire tube. 

To understand how each cavity component contribute to the overall losses and in turn influence the quality factor in this type of resonator, we have conducted FEM simulations taking into account the temperature dependence of losses in sapphire and copper.    
We have then measured $Q_0$ and $f$ for several values of temperature in the range (4-293)\,K, and compared the experimental data with results from the simulations. 

In table \ref{tab:1} we report the simulations parameters and the results, in which we decouple the losses in copper and sapphire, together with the corresponding experimental data. 
\begin{table*}[t]
\begin{center}
 \renewcommand\arraystretch{1.3} 
\begin{tabularx}{\textwidth}{Y| Y Y Y | Y Y Y Y Y | Y}
\toprule
T (K) & $\sigma_{Cu}^{\rm eff}$ & $\epsilon_{r}$ & $\tan\delta$ & f (GHz) & $C_{030}$ & $Q_{Sapph}$ & $Q_{Cu}$ & $Q_{0}$ & $Q_{0,exp}$  \\ [0.5ex] 
\midrule
293 & 37.75	 & 11.44 & 4.87$\cdot10^{-6}$ & 10.4468 & 0.0321 & 1.21$\cdot10^{6}$ & 2.04$\cdot10^{6}$ & 7.60$\cdot10^{5}$ & 6.17$\cdot10^{5}$\\

260 & 40.55 & 11.3889 & 3.93$\cdot10^{-6}$ & 10.4529 & 0.0321 & 1.50$\cdot10^{6}$ & 2.18$\cdot10^{6}$ & 8.88$\cdot10^{5}$ & 7.28$\cdot10^{5}$ \\

230 & 43.70 & 11.3486 & 2.88$\cdot10^{-6}$ & 10.4576 & 0.0321 & 2.05$\cdot10^{6}$ & 2.35$\cdot10^{6}$ & 1.09$\cdot10^{6}$ & 8.08$\cdot10^{5}$ \\

190 & 49.34 & 11.2953 & 1.81$\cdot10^{-6}$ & 10.4638 & 0.0320 & 3.26$\cdot10^{6}$ & 2.64$\cdot10^{6}$ & 1.46$\cdot10^{6}$ & 1.09$\cdot10^{6}$  \\

160 & 55.53 & 11.2602 & 7.84$\cdot10^{-7}$ & 10.4676 & 0.0320 & 7.51$\cdot10^{6}$ & 3.02$\cdot10^{6}$ & 2.12$\cdot10^{6}$ & 1.48$\cdot10^{6}$  \\

125 & 67.57 & 11.2288 & 2.56$\cdot10^{-7}$ & 10.4711 & 0.0320 & 2.29$\cdot10^{7}$ & 3.58$\cdot10^{6}$ & 3.09$\cdot10^{6}$ & 2.19$\cdot10^{6}$  \\

100 & 84.17 & 11.2125 & 8.87$\cdot10^{-8}$ & 10.4729 & 0.0320 & 6.62$\cdot10^{7}$ & 4.47$\cdot10^{6}$ & 4.15$\cdot10^{6}$ & 2.80$\cdot10^{6}$  \\

77 & 114.73 & 11.2044 & 2.48$\cdot10^{-8}$ & 10.4738 & 0.0320 & 2.37$\cdot10^{8}$ & 6.10$\cdot10^{6}$ & 5.88$\cdot10^{6}$ & 3.33$\cdot10^{6}$  \\

50 & 174.57 & 11.2006 & 7.52$\cdot10^{-9}$ & 10.4744 & 0.0320 & 7.82$\cdot10^{8}$ & 9.29$\cdot10^{6}$ & 9.02$\cdot10^{6}$ & 4.63$\cdot10^{6}$  \\

25 & 205.16 & 11.2002 & 5.34$\cdot10^{-9}$ & 10.4745 & 0.0321 & 1.10$\cdot10^{9}$ & 1.09$\cdot10^{7}$ & 1.06$\cdot10^{7}$ & 5.48$\cdot10^{6}$  \\

4.2 & 207.04 & 11.2 & 1.19$\cdot10^{-9}$ & 10.4745 & 0.0321 & 1.03$\cdot10^{9}$ & 1.10$\cdot10^{7}$ & 1.08$\cdot10^{7}$ & 6.23$\cdot10^{6}$ \\
\bottomrule
\end{tabularx}
  \caption{Parameters set in the FEM simulations and results. Values are given at different temperatures for the effective copper  conductivity $\sigma_{Cu}$, the relative dielectric constant $\epsilon_r$ and the tangent loss of sapphire. Losses in copper ($Q_{\rm Cu}$) and in sapphire ($Q_{Sapph}$) are quantified by respectively running a model in which $\tan \delta$ is set to null value, and another model with lossless cavity walls. 
  In the last column we report for comparison the measured unloaded quality factor $Q_{0,exp}$.}
  \label{tab:1}
\end{center}
\end{table*}

The losses are additive, and their effect on the quality factor can be estimated through the simulations according to
\begin{equation}
    \frac{1}{Q_{0}}=\frac{1}{Q_{\rm Cu}}+\frac{1}{Q_{Sapph}} = R_s/G+\tan \delta \; ,
\end{equation}
where $G$ is a constant that can be calculated depending on the cavity geometry and the chosen mode, $R_s$ is the copper surface resistance and $\tan \delta$ is the tangent loss of sapphire. Here, $Q_{\rm Cu}$ and $Q_{Sapph}$ are the quality factors of the modeled cavities with losses taking place entirely in copper or in sapphire, respectively. 

The loss tangent at fixed temperature can significantly vary depending on crystallographic orientation, and the concentration of impurities or dislocations. A tangent loss value as low as $10^{-10}$ has been reported in sapphire at cryogenic temperatures \cite{Buckley:1994}. In the simulations we used the values reported in table \ref{tab:1}, taken from Ref.\,\onlinecite{braginsky1987experimental}, corrected to account for the different frequency \cite{shtin2009theory}.

In Fig.\,\ref{fig:1}\,(d) we show the surface loss profile of the TM$_{030}$ in the copper walls, obtained running the FEM cavity model at 4.2\,K. 
Cone structures at the end caps are used to reduce current dissipation, a critical aspect when highest quality factors are targeted. For instance, dissipation in the endcaps (shown in Fig.\,\ref{fig:1} (e)) for the mode of interest is limited to 16$\%$ of the overall dissipation due to copper finite conductivity. 

As concerns the teflon screws, their contribution to overall cavity loss can be neglected. In fact, by attributing a value of $\tan\delta=10^{-3}$ to teflon at room temperature the value $Q_0=5.8\times 10^{8}$ is obtained in the modeled cavity with lossless metallic walls and dielectric cylinders.

The values of surface effective conductivity of copper $\sigma_{Cu}^{eff}$  used in the model have been computed at 10.4\,GHz in the anomalous skin depth regime \cite{Calatroni:2718002}.

For each temperature we account for the changes of cavity geometry using the thermal expansion coefficients of copper \cite{simon1992nist} and sapphire \cite{bradley2013properties}.
We also considered the temperature dependence of the dielectric permittivity $\epsilon_r$ as reported in Table \ref{tab:1}, where the values in the third column have been calculated starting from the value we measured at 4.2\,K ($\epsilon_r=11.2$) and rescaling the data reported in Ref.\,\onlinecite{krupka1999complex}. 
In table \ref{tab:2} we report simulation results for the modeled cavity at 4\,K with a few values of $\epsilon_r$ including the values expected at 293\,K and 4.2\,K.   
As can be inferred by comparison of the simulation results in Table \ref{tab:1} and in Table \ref{tab:2}, we can ascribe a 19\,MHz frequency shift when the cavity is cooled from room temperature to 4.2\,K. 
\begin{table}
\begin{center}
 \renewcommand\arraystretch{1.3} 
\begin{tabularx}{8.5cm}{Y Y Y Y Y Y}
\toprule
$\epsilon_{r}$ & f (GHz) & $C_{030}$ & $Q_{Cu}$ & $Q_{Sapph}$ & $Q_{0}$ \\ [0.5ex] 
\midrule
11.44 & 10.4556 & 0.0324 & 1.14$\cdot10^{7}$ & 4.98$\cdot10^{9}$  & 1.11$\cdot10^{7}$  \\ 
11.3 & 10.4666 & 0.0322 & 1.10$\cdot10^{7}$ & 4.96$\cdot10^{9}$  & 1.10$\cdot10^{7}$  \\
11.2 & 10.4745 & 0.0321 & 1.09$\cdot10^{7}$ & 1.03$\cdot10^{9}$  & 1.08$\cdot10^{7}$ \\
11.1 & 10.4825 & 0.0319 & 1.07$\cdot10^{7}$ & 1.21$\cdot10^{9}$  & 1.06$\cdot10^{7}$ \\
\bottomrule
\end{tabularx}
  \caption{
  Simulation of the cavity at 4.2\,K for different values of $\epsilon_r$. The cavity frequency changes by 19\,MHz (relative change ) using the values reported in table \ref{tab:1} at room temperature and 4.2\,K, while $Q_0$ and $f$ are not significantly influenced. }
  \label{tab:2}
\end{center}
\end{table}
An additional shift of 9 MHz is instead determined by variation of the cavity geometry due to thermal expansion of the copper and sapphire cylinders. 
Moreover, simulation results in Table \,\ref{tab:2} show that the quality factor is not significantly affected by $\epsilon_r$ temperature changes.

This resonator is expected to have a quality factor as high as $11\times 10^6$ at liquid helium temperature, calculated with parameters $\tan \delta= 1.19\times 10^{-9}$, $\epsilon_r=11.2$ and surface copper resistivity $0.00483\,\Omega$.
The form factor, which can only be calculated through FEM modeling, is fixed to $C_{030}=$0.032 as shown in table \ref{tab:1} and \ref{tab:2}.
The $C_{030} V$ product is thus $ 34.6$ \,cm$^3$, compared to 164.6\,cm$^{3}$ for the TM$_{010}$ mode of an empty cylindrical cavity with the same length resonating at 10.4\,GHz. 

\begin{figure}[h!]
\includegraphics[width=3.2in]{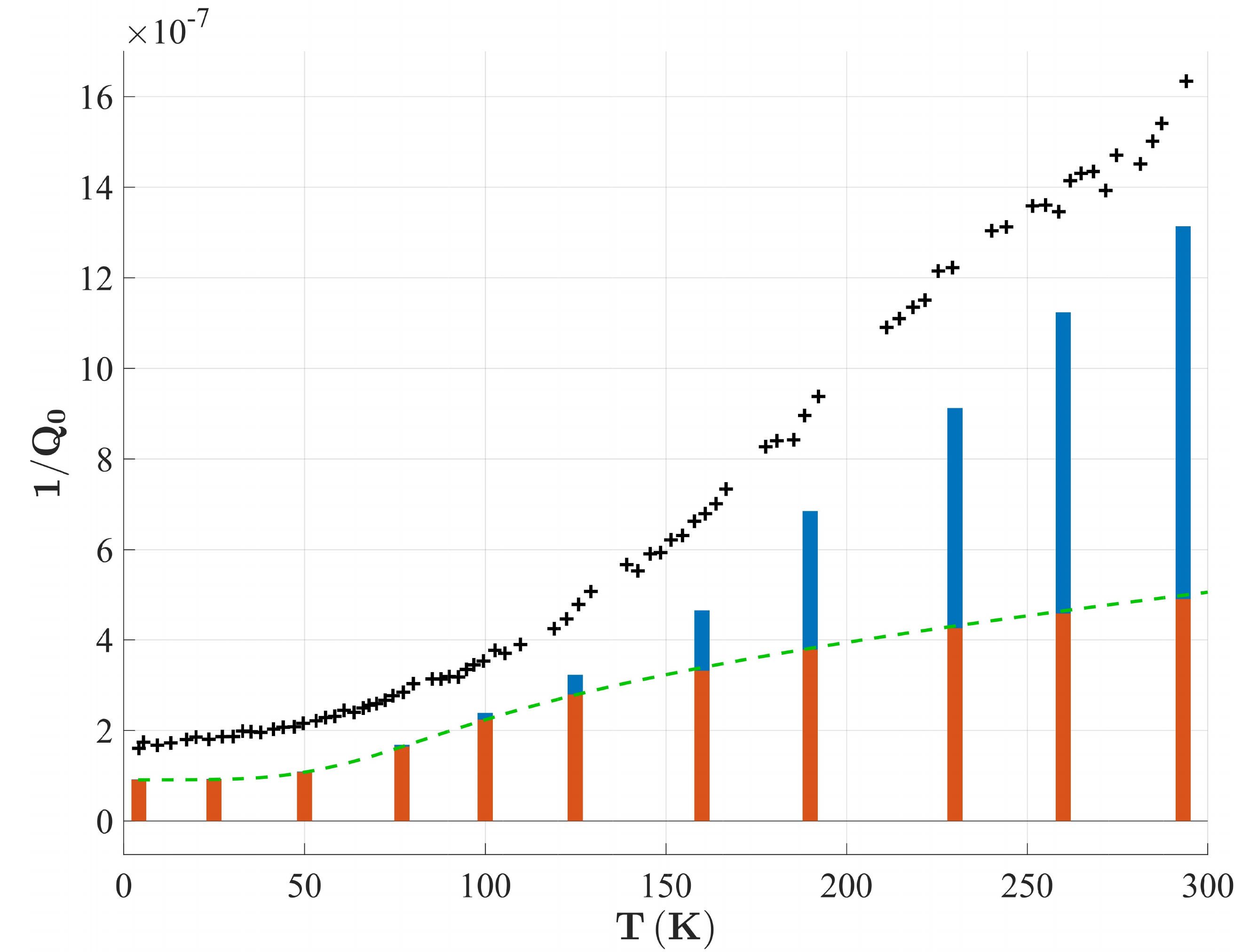}
\caption{\label{fig:3a} Cavity dissipation according to the simulation. Loss in copper is represented in orange, that for sapphire is in blue. For comparison, we report in the same plot experimental data. 
}
\end{figure}

In Fig.\,2 the simulation results show that at room temperature the losses in sapphire exceed those in copper, but they quickly decrease due to the loss tangent rapid drop. 
The quality factor of the modeled cavity saturates to 11 millions once below about 25\,K, as at this point the copper conductivity does not vary significantly.

\section{\label{sec:measurements} Measurements in the (4.2-293)\,K temperature range}

 We measure the resonant frequency of the TM$_{030}$ mode and its quality factor by recording at a Vector Network Analyzer (VNA) the $S_{21}$ scattering parameter of two coaxial antennas weakly coupled to the cavity mode.  
 
\begin{figure}[h!]
\includegraphics[width=3.3in]{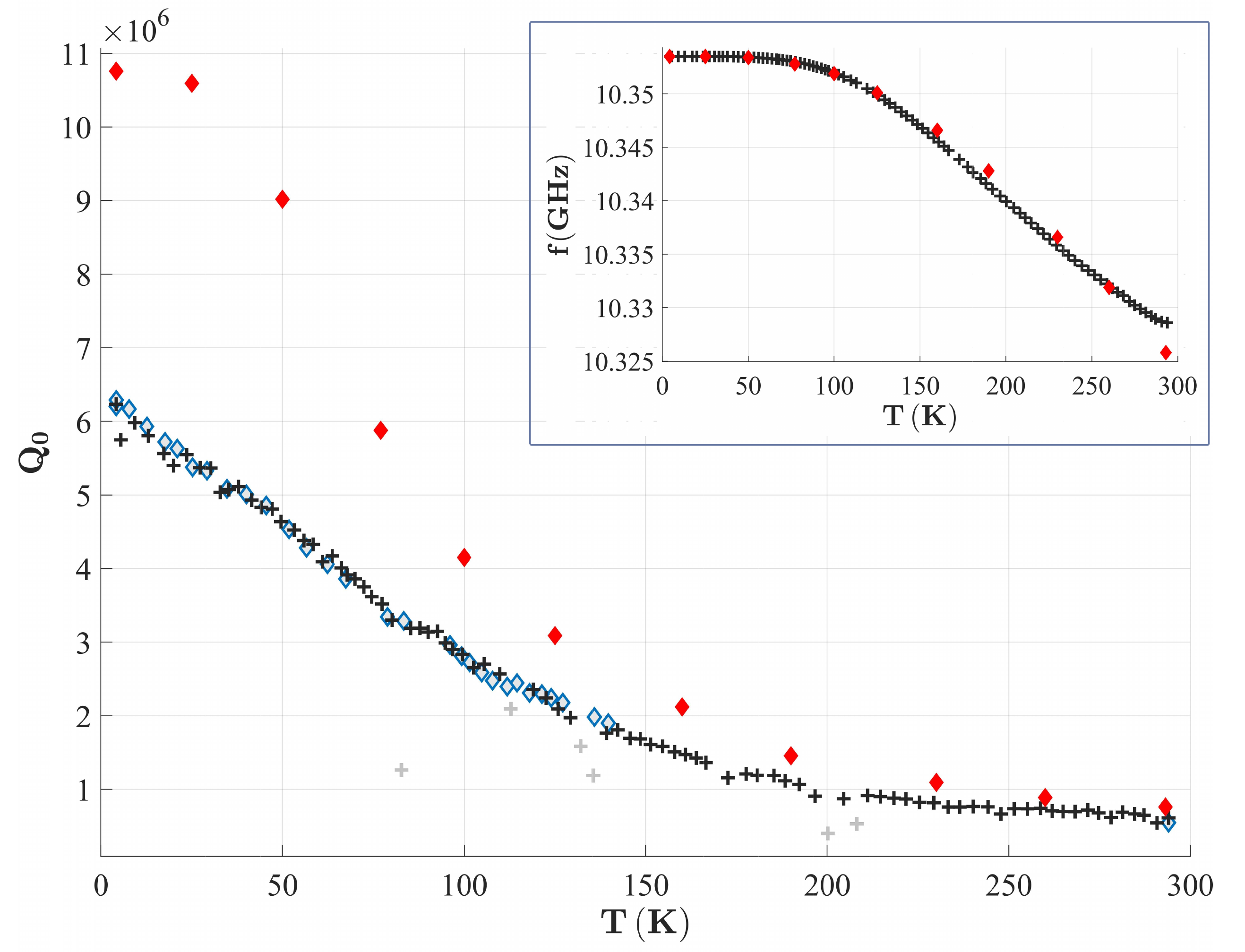}
\caption{\label{fig:3} Temperature dependence of the unloaded cavity quality factor. The light blue circles and black cross correspond to data series acquired during two different cooling runs. Red diamond data points are the values of quality factors obtained from the simulation.
Inset: Temperature dependence of the TM$_{030}$ mode frequency. Note that the resonant frequency values obtained from the simulations have been shifted by 121\,MHz to match the value measured at 4\,K.
}
\end{figure}

 The cavity can be cooled to $\sim $4\,K within a vacuum chamber designed to allow operation inside cryogenic dewars, and a thermometer calibrated to within 1\,K has been used to measure the temperature. To ensure proper cooling of the cavity components, the chamber is loaded with He exchange gas.
 Fig.\,\ref{fig:3} shows the unloaded quality factor measured at different temperatures in the range (4.2-293)\,K, together with simulations results.  Deviations from the data trend are observed when the TM$_{030}$ mode mixes with other TE or TEM cavity modes. 

While discrepancy between experimental data and the simulation results is within 20\,\% at room temperature, we observe in Fig.\,\ref{fig:3} an increasing deviation when the cavity is cooled down, suggesting that the dielectric losses in sapphire play a role also at low temperature. 
As shown in the inset of Fig.\,\ref{fig:3} the overall frequency shift of about 28\,MHz predicted by the simulations for the temperature change from 293\,K to 4.2\,K, is in good agreement with the experimental results.

\section{Measurements at liquid helium temperature in high magnetic fields}\label{mag}

\begin{figure}[h!]
\includegraphics[width=3.3in]{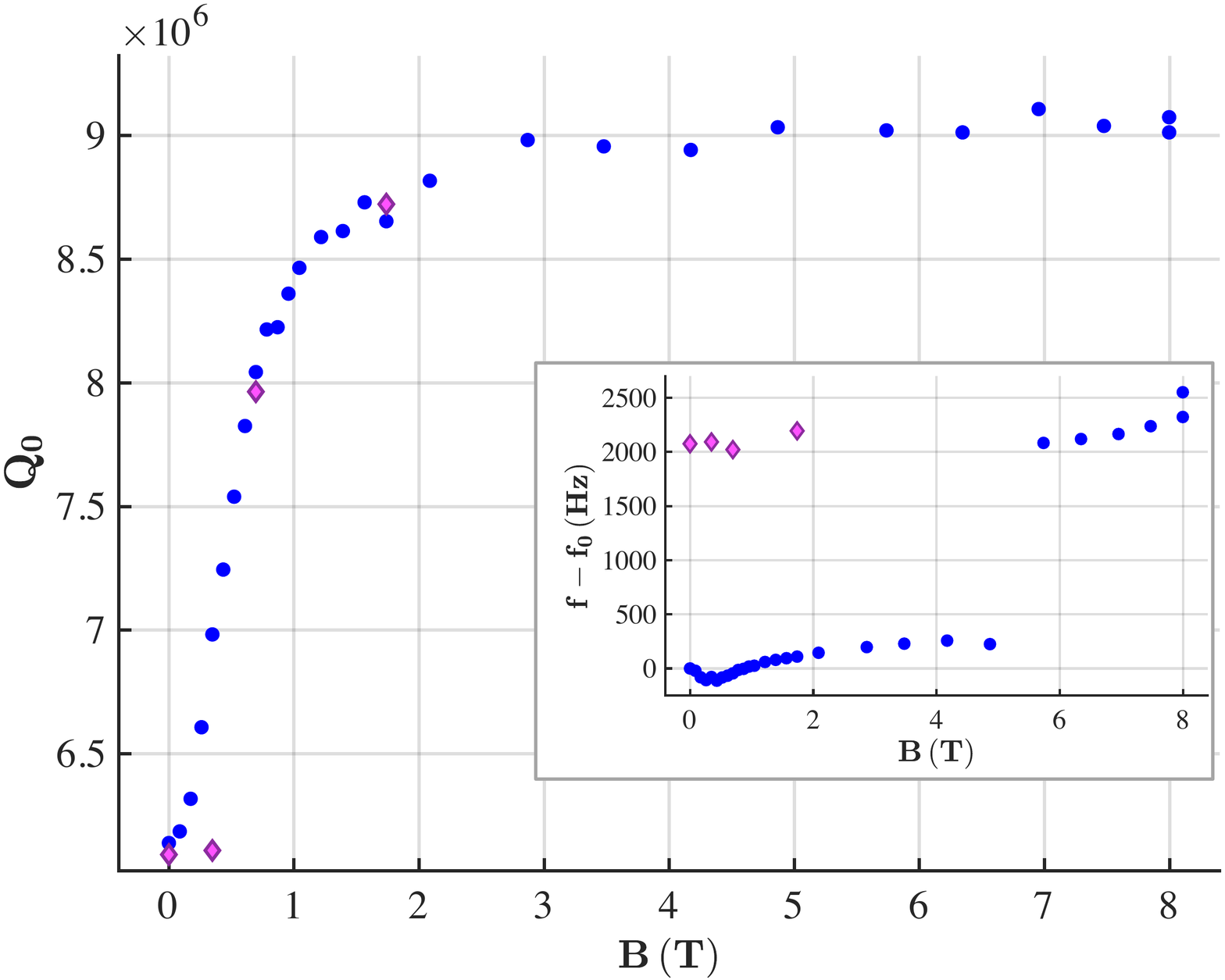}
\caption{\label{fig:4} Unloaded quality factor vs applied magnetic field. Diamond data points were acquired during magnetic field ramp down.
Inset: Corresponding frequency shift. Note that at about 5\,T the cavity frequency shifts by about 1.7\,kHz, value comparable to the unloaded cavity linewidth $\Delta f = f/Q_0 \simeq$\,1.1\,kHz.}
\end{figure}

As this resonator is targeted to axion detection, it is important to measure its parameters under intense magnetic fields.
Bulk copper undergoes magnetoresistance \cite{DELAUNAY195937}, but as shown in recent studies \cite{Ahn_2017}, the fractional change in quality factor of copper resonators immersed in B fields between 6 to 10\,T is always positive, of about 2\,\% percent. 
Fig.~\ref{fig:4} shows the results of measurements with the present dielectric cavity inserted inside the 150 mm-diameter bore of an 8.1 T superconducting (SC) magnet of length 500 mm. 

The magnet, manufactured by Cryogenic Ltd \cite{cryogenic}, was cooled in LHe and can be ramped up to a maximum of 8\,T over less than 1 hour. To record the quality factor vs B-field dependence, we manually set current values using smaller steps between 0 and 2\,T, and  larger ones afterwards.
Different regimes can be identified in the measured quality factor vs B-field plot: a rapid increase up to 0.8\,T, a knee between 0.8 and 2\,T followed by a small growth up to approximately 5\,T. For values exceeding 5\,T, a quality factor of about 9\,millions is recorded, $50\%$ greater than the value at null field.
Remarkably, the measured maximum value at 8\,T is only about $20\%$ lower than the value expected from the simulations.
The unexpected increase of quality factor in the presence of intense magnetic fields can be attributed to the magnetic properties of the dielectric cylinders. In fact, high purity sapphire crystals are known to host a number of paramagnetic impurities \cite{Farr:2013,Giordano:2020} as Cr$^{3+}$, Fe$^{3+}$ and Mo$^{3+}$. V$^{2+}$ spin ensembles are also expected if the crystal has been subjected to radiation treatment or annealing procedures\cite{Farr:2013} to relax internal stresses. Note that the present sapphire tubes have been annealed in vacuum at a temperature of 1900\,C degrees.

Even when their concentration is as low as parts per billion, impurities can be detected and investigated by identifying avoided crossings between their electron spin resonances (ESR) and photonic whispering gallery modes \cite{Farr:2013,Giordano:2020}. 
Measured properties are the zero field splitting, and the Landé $g$ factors are obtained through the gradient of the frequency dependence on magnetic field of detected ESR transitions $df/dB=g\mu_B/\hbar$, with $\mu_B$ being the Bohr magneton. 
For instance, reported spin transition lines at zero field for Fe$^{3+}$, Cr$^{3+}$ are at 12.03\,GHz and $11.45$\,GHz respectively, while 8 lines in the range $8.7-10.4$\,GHz are attributed to the ion V$^{2+}$. With the application of a magnetic field, these absorption lines are tuned away from the TM$_{030}$ cavity mode at the rate $\sim 28$\,GHz/T for Cr and V impurities, and even faster for some transitions in Fe \cite{Farr:2013}.
Therefore we might expect that already at a few Tesla magnetic field amplitude, the related dissipative channel is significantly suppressed, giving a plausible explanation of the observed $Q_0$ vs $B$ data trend.  


\section{tuning}\label{tuning}

\begin{figure}
\includegraphics[width=3.3in]{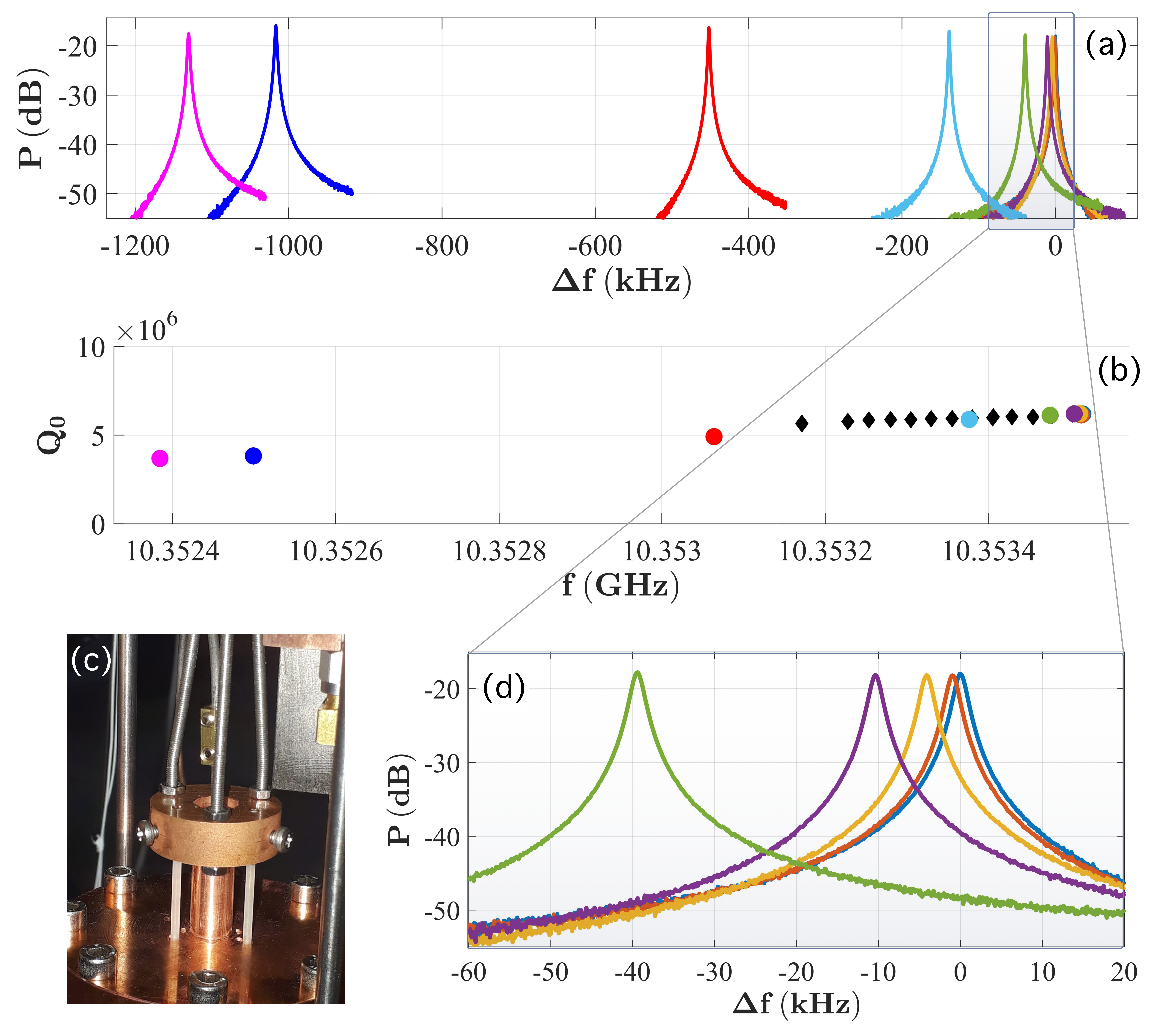}
\caption{\label{fig:5} Cryogenic frequency tuning. (a)  
Maximum 1.2\,MHz frequency shift obtained and (b) measured $Q_0$ values exceeding $Q_a$. (c) Picture of the rods positioning system (top). Tuning is accomplished by displacing triplets of 2\,mm-diameter sapphire rods relative to the top and bottom cavity endcaps. The triplets are anchored to copper blocks hanging on steel wires, that allow for controlling the rods position inside the cavity with an external micrometer. (d) Fine tuning. 1\,kHz is the shown minimum frequency shift.}
\end{figure}

To make this resonator suitable for axion searches, the internal sapphire tube is substituted by two hollow half cylinders that allow for tuning the TM$_{030}$ frequency up to about 500\,MHz when moved apart along the radial direction \cite{Kim:2019asb,Alesini:2020vwh}. To demonstrate such a wide tuning range, a thorough study is under way involving simulations and in situ tests, here we report a tuning method that was implemented to tune over a narrow range the same cavity used in the measurements presented in previous sections. The objective in this case is to investigate the experimental requirements that are needed for running the experiment with a cavity exceeding the axion quality factor $Q_a$. In this study, that is currently ongoing in our laboratory, capability to shift the cavity central frequency within a fraction of its bandwidth is required. 
By displacing triplets of 2\,mm-diameter sapphire rods relative to the top and bottom cavity endcaps, we shift the TM$_{030}$ mode frequency without impacting the quality factor as shown in Fig.\,\ref{fig:5}. 
The maximum shift we obtain is of about $1.2\,$MHz, with 1\,kHz minimum frequency change, comparable with the cavity bandwidth.  

\section{\label{sec:disc} Conclusions}

Microwave cavity resonators employed in axion dark matter search shall exhibit high quality factors in the presence of multi-tesla fields. In this work we have demonstrated the highest quality factor ever reported in the literature for a microwave cavity under intense magnetic fields, reaching more than 9\,millions at 8\,T. This result has been obtained with a copper cavity hosting two dielectric hollow cylinders, a design inspired by the idea that higher order modes in cavity resonators can effectively be used to address the high-frequency challenge in haloscope search when a low loss dielectric material is used to properly shape the electric field of the chosen mode.  
Practical limitations related to operating such a complex geometry at cryogenic temperatures \cite{Alesini:2020vwh} have been overcome in the present design, which has been perfected via comparison of the results of finite element modeling and tests in the laboratory. Measurements of the unloaded quality factor and resonance frequency of the TM$_{030}$ axion-sensitive mode have been conducted at various temperatures in the range $(4.2-293)$\,K, to determine the interplay between dissipation mechanisms in the copper walls and in the sapphire tubes. At 4.2\,K we measured $Q_0=6.23\times 10^6$, apparently limited by the copper finite conductivity as calculated taking into account the anomalous skin effect. 
Unexpectedly, the $Q_0$ measured under an 8\,T-field exceeds the value measured at null field by about 50\%. This experimental result is close to the values obtained from the simulations, in which we assumed that the tangent loss in sapphire decreases down to about $10^{-9}$ at 4.2\,K, in agreement with several works reported in the literature.

As the losses in cooper via magnetoresistance are expected to be within a few percent \cite{Ahn_2017}, in the measurements at low B-fields we identified a posteriori an additional loss channel due the magnetic susceptibility of sapphire. Paramagnetic impurities, previously investigated in high purity sapphire crystals, have indeed several zero point field transitions around the TM$_{030}$ frequency, that change at the rate $df/dB\sim 28$\,GHz/T. Therefore at the high fields required for axion detection they are completely swept away from the TM$_{030}$ frequency. 

As losses in copper can be accurately predicted, and having suppressed paramagnetic losses in sapphire at high magnetic field, we can obtain an estimate of the $\tan\delta$ of the sapphire tubes from the high field experimental results. 
The measured quality factor at 4.2\,K is obtained in the FEM simulation if a tangent loss of $2.0\times 10^{-8}$ is assumed in place of $1.2\times 10^{-9}$.   

Our cavity development has involved precision metrology and modeling in simulation environments such as HFSS of ANSYS Electronics \cite{hfss}, demonstrating the capability to improve our understanding of this type of dielectric resonator and to optimize its future designs and tests that will include the wider tunability characterization.
Across this study, measured and simulated results are in rather good agreement. Small quantitative differences are found within $\sim$1\%, which is expected, given the complexity of the simulated model and taking into account mechanical tolerances.

We can use the scan rate expression given by eq.\,\ref{eq:3} to ponder the potential of this resonator for axion DM search with haloscopes. If $\beta_{opt}$ is the value of coupling coefficient that maximizes the scan rate for a given $Q_c/Q_a$ ratio \cite{Kim:2020kfo}, we estimate about $\sim 15\,$MHz/day for $\beta_{opt}\sim 7$ when a goal sensitivity to KSVZ family of models ($g_{\gamma}=-0.97$) at 95\% confidence level ($\Sigma=2$) is addressed, assuming quantum-limited readout and B$_0=14$\,T. 
A reasonable signal integration time of about 50\,s would then be needed for a single measurement. 
For comparison, a conventional copper cavity resonating in its TM$_{010}$ at the same frequency ($r=0.011\,$m, C$_{010}=0.5$, $Q_0\simeq20000$) would reach the same sensitivity at $\sim 0.35\,$MHz/day at $\beta_{opt}\sim 4$, for which a feasible search is questionable.

\begin{acknowledgments}
This material is based upon work supported by INFN (QUAX experiment) and the U.S. Department of Energy, Office of Science, National Quantum Information Science Research Centers, Superconducting Quantum Materials and Systems Center (SQMS) under the contract No. DE-AC02-07CH11359. \\
We are grateful to E. Berto (University of Padova and INFN) who substantially contributed to the mechanical realization of this cavity and of its tuning system. A. Benato (INFN) did part of the mechanical work and M. Zago (INFN) the mechanical drawings. 
The contribution of F. Calaon and M. Tessaro (INFN) to the experiment cryogenics and electronics is gratefully acknowledged.
The cryogenic service of the Laboratori Nazionali di Legnaro provided the liquid helium to run the described experiments. 

\end{acknowledgments}

\nocite{*}
\bibliography{Q10M}

\end{document}